\documentclass[showpacs,preprintnumbers,twocolumn,amsmath,amssymb]{revtex4}
\usepackage{graphicx}
\usepackage{dcolumn}
\begin{document}

\title{Spontaneously induced atom-radiation entanglement in an ensemble of two-level atoms}
\author{Sintayehu Tesfa}
\email{sint_tesfa@yahoo.com} \affiliation{Physics Department,
Addis Ababa University, P. O. Box 1176, Addis Ababa, Ethiopia}

\date{\today}

\begin{abstract}Analysis of the spontaneously induced correlation on atom-radiation
entanglement in an ensemble of two-level atoms initially prepared
in the upper level and placed in a cavity containing a squeezed
radiation employing the method of evaluating the coherent-state
propagator is presented. It is found that the cavity radiation
exhibits squeezing which is directly attributed to the squeezed
radiation in the cavity. The intensity of the cavity radiation
increases with the squeeze parameter and interaction time. It is
also shown that substantial degree of entanglement between the
atomic states and radiation mode exits at a particular time that
depends on the coupling constant and squeeze parameter. We come to
understand that though the squeezed radiation is directly
accountable for the cavity squeezing, it significantly destroys
the atom-radiation entanglement.
\end{abstract}
 \maketitle

\section{INTRODUCTION}

Interaction of an electromagnetic radiation with atoms is one of
the oldest \cite{zphys31681} and central problems in quantum
optics. In particular, the interaction of several two-level atoms
\cite{pra65053819, prl831319, prl772658,
prl803948,jmo502765,sint,pra211297,pra423051} as well as a single
two-level atom \cite{jmo451859} with a single-mode squeezed
radiation have attracted the attention of several authors over the
years. In addition to the existing huge theoretical
investigations, there are also significant amount of experimental
works related to two-level atoms in an optical cavity pumped
externally with a coherent radiation, wherein the coupling through
the mirrors is taken into consideration
\cite{prl831319,prl803948}. Earlier, Orozco {\it{et al.}}
\cite{josab41490} carried out an experiment to study the
nonclassical properties of the radiation in an optical cavity that
contains N two-level atoms in a strong coupling regime, where the
atomic spontaneous emission rate into modes other than the
resonant cavity mode is much greater than the damping rate of the
cavity radiation, and found that the generated light exhibits
squeezing. Most recently, entanglement of atoms with radiation has
been considered by various authors \cite{prl96030404,pra70043809}
and encouraging results have been obtained.

Nowadays, correlated quantum systems of atoms or radiations or
atom(s) with a radiation have gained a considerable importance in
connection with the potential role they are envisioned to play in
the fields of quantum information and measurements beyond quantum
limit. In the quantum information theory, the issue of mapping the
quantum states of the carrier (light) onto the quantum states of
the storing device (atoms) is one of the essential problems that
are not completely resolved at present. The mechanism of employing
atomic ensembles to momentarily store quantum states of light via
entanglement can serve as a device for interfacing the photonic
quantum memories with processors. Atom-radiation entanglement is
not only crucial for many applications in the long range quantum
communications, but also it is the key element to give the final
answer to the Einestein-Podolsky-Rosen's \cite{pr47777} question
on the real properties of nature, which they argued that there is
inconsistencies between quantum theories and their idea of local
and deterministic descriptions of nature \cite{prl91110405}. In
this regard, there is a proposal that involves mapping of a state
of propagating quantum correlated light onto an atomic ensemble,
which is relevant for the storage of the entanglement in the state
of the light \cite{prl831319}. Moreover, as recently shown by
Andre and Lukin \cite{pra65053819}, based on the analogy of the
optical parametric oscillation to the process that transfers pairs
of atoms from the lower level to a well defined final upper level,
 there is quantum correlations among the atomic states.

In relation to the importance of the issue, there are abundant
works in the literature that address the possible mechanisms for
generating entanglement in radiation modes
\cite{pra74023816,sintpra,pra74} and in an ensemble of atomic
systems \cite{prl831319}. Besides, there is an evidence which
indicates that the entanglement between atoms in the same ensemble
is potentially applicable in the precession measurements
\cite{prl793865}. The interaction among atoms through exchange of
photons is found to be responsible for the entanglement and spin
squeezing that are characterized by the reduced fluctuations in an
observable with an increased fluctuations in the canonical
conjugate just like entanglement and squeezing in the radiation
\cite{pra475138}. On the other hand, as thoroughly discussed by
Walser {\it{et al.}} \cite{prl772658} the method of mapping the
state of a single quantized cavity mode adiabatically onto a
finite dimensional sub-manifold of a two-level atom that passes
through the resonator can be formulated using angular momentum
operator. But, the first time analysis and observation of
entanglement between the polarization of a single photon and the
internal state of a single neutral atom stored in an optical
dipole trap during the spontaneous emission is reported by Votz
{\it{et al.}} \cite{prl96030404}.

In this contribution, we seek to investigate the mechanism that
establishes the atom-radiation entanglement analytically. To
achieve our goal, we consider the correlation created between
atoms and cavity radiation due to the spontaneous emission from
the two-level atoms initially prepared in the upper level and then
placed in a cavity, which is called as spontaneously induced
atom-radiation entanglement. We take the cavity to be of a
nonleaky type, since the present state-of-the-art cavity QED
experiments are very close to such a limit \cite{jpbamop38551}.
One may expect that the strong correlation between the modes in
the squeezed radiation can create additional correlation between
the atomic states and radiation in the cavity, and hence leads to
enhancement of atom-radiation entanglement. In connection to this,
Hald {\it{et al.}} \cite{prl831319} demonstrated the possibility
for mapping a quantum state of free propagating squeezed light
onto a multi-atom ensemble as proposed by Kuzmich {\it{et al.}}
\cite{prl794782} applying the spin polarization of Cesium atoms in
excited state. In accordance to this, we investigate the
entanglement between the quantum state of an ensemble of two-level
atoms with the cavity radiation using the Schweninger's
representation of the angular momentum in terms of the boson
operators employing the approach recently introduced \cite{pra74}
based on the criterion set by Duan {\it{et al.}} \cite{duan}. We
also study the photon statistics of the cavity as well as the
emitted radiation. We, in particular, evaluate the quadrature
variances of the cavity radiation, mean photon number available in
the cavity, mean photon number emitted by the atoms, amount of the
atom-radiation entanglement, and photon number distribution of the
cavity radiation.

\section{Q-function}

The Hamiltonian describing the interaction of two-level atoms with
a single-mode cavity radiation has been derived by several authors
\cite{pra211573,sint} based on Jaynes-Cummings model. As clearly
indicated, for example, in the work of Brecha {\it{et al.}}
\cite{pra592392} this interaction Hamiltonian, in the
rotating-wave and electric-dipole approximations, can be expressed
as
\begin{align}\label{nt01}\hat{H}_{I}=ig\left[\hat{a}^{\dagger}\hat{J}_{-}
-\hat{a}\hat{J}_{+}\right],\end{align} where $g$ is the coupling
constant that measures the strength of the interaction between the
radiation and atoms, $\hat{a}$ represents the cavity radiation,
whereas
\newline$\hat{j}_{\pm}=\sum_{j=1}^{N}\hat{\sigma}_{\pm}^{j}e^{\pm
i\vec{k}.\vec{r}_{j}}$ are the collective atomic operators
\cite{pra423051} that are expressible in Schweninger's
representation of the angular momentum operators in terms of the
usual boson operators as
\begin{align}\label{nt02}\hat{J}_{+}=\hat{b}^{\dagger}\hat{c},\end{align}
\begin{align}\label{nt03}\hat{J}_{-}=\hat{c}^{\dagger}\hat{b},\end{align}
where $\hat{b}$ and $\hat{c}$ correspond to the atoms in the lower
and upper levels. Then, making use of Eqs. \eqref{nt02} and
\eqref{nt03} the interaction of two-level atoms with a single-mode
cavity radiation can be described by the quantum Hamiltonian of
the form
\begin{align}\label{nt04}\hat{H}_{I}=ig\left[\hat{a}^{\dagger}\hat{b}^{\dagger}\hat{c}
-\hat{a}\hat{b}\hat{c}^{\dagger}\right].\end{align} It is a
well-known fact that the differential equations following from a
trilinear Hamiltonian are nonlinear and hence difficult to
analytically solve directly. As a result, in present work we
consider the case when almost all the atoms are prepared to be
initially in the upper level to overcome this problem. In this
case, it is justifiable to treat the operator $\hat{c}$ as a
c-number $\gamma_{0}$ which is taken to be real-positive constant
\cite{pra211297}. Therefore, in this approximation the Hamiltonian
\eqref{nt04} reduces to
\begin{align}\label{nt05}\hat{H}_{I}=i\lambda\left[\hat{a}^{\dagger}\hat{b}^{\dagger}
-\hat{a}\hat{b}\right],\end{align} in which $\lambda=g\gamma_{0}$.
In principle, it is possible to study the quantum features of such
an interaction by solving the differential equations following
from the pertinent Heisenberg or quantum Langevin equations.
However, in this contribution we wish to exploit the advantageous
conferred by the quasi-probability distribution functions.

It is a well-established fact that the Q-function can be defined
for a two-mode cavity radiation initially in squeezed state as
\begin{align}\label{nt06}Q(\alpha,\beta,t)={1\over\pi^{2}}\int\langle\alpha,\beta|\hat{\rho}(t)
|\alpha,\beta\rangle,\end{align} where
\begin{align}\label{nt07}\hat{\rho}(t)=\hat{U}(t)|r,0\rangle\langle0,r|\hat{U}^{\dagger}(t),\end{align}
in which $\hat{U}(t)=\exp(-i\hat{H}_{I}t)$ is the evolution
operator corresponding to the interaction Hamiltonian \eqref{nt05}
and $|r,0\rangle$ is the initial state of the cavity radiation,
where $r$ is the squeeze parameter. Upon introducing the two-mode
completeness relation for coherent states,
\newline$\hat{I}=\int{d^{2}\gamma\over\pi}{d^{2}\eta\over\pi}|\gamma,\eta\rangle\langle\eta,\gamma|$,
and the fact that \cite{sint}
\begin{align}\label{nt08}\langle\gamma|r\rangle=\sqrt{{1\over\cosh r}}\exp\left[
-{\gamma^{*}\gamma\over2}-{\gamma^{*^{2}}\over2}\tanh r
\right],\end{align} we come up with
\begin{align}\label{nt09}Q(\alpha,\beta,t)&={1\over\pi^{2}\cosh r}\int
{d^{2}\gamma\over\pi}{d^{2}\eta\over\pi}{d^{2}\mu\over\pi}{d^{2}\nu\over\pi}
\notag\\&\times K(\alpha,\beta,t|\gamma,\eta,0)
K^{*}(\alpha,\beta,t|\nu,\mu,0)\notag\\&\times\exp\left[-{1\over2}\big(\gamma^{*}\gamma
+\eta^{*}\eta+\nu^{*}\nu+\mu^{*}\mu\right.\notag\\&\left.+\big(\gamma^{*^{2}}+\mu^{2}\big)\tanh
r \big)\right],\end{align} where
\begin{align}\label{nt10}K(\alpha,\beta,t|\gamma,\eta,0)=\langle\alpha,\beta|
\hat{U}(t)|\gamma,\eta\rangle\end{align} is the coherent-state
propagator for a two-mode radiation.

Following the method introduced in Ref. \cite{pra465379} and
latter employed in studying degenerate parametric oscillator
\cite{oc151384}, the coherent-state propagator pertinent to the
Hamiltonian \eqref{nt05} is found to be
\begin{align}\label{nt11}K(\alpha,\beta,t|\gamma,\eta,0)&={1\over\cosh\lambda t}\exp
\left[-{\alpha^{*}\alpha\over2}-{\beta^{*}\beta\over2}-{\gamma^{*}\gamma\over2}
\right.\notag\\&\left.-{\eta^{*}\eta\over2}+\big(\alpha^{*}\beta^{*}-\gamma\eta\big)
\tanh\lambda t\right.\notag\\&\left.+{1\over\cosh\lambda
t}\left(\alpha^{*}\gamma+\beta^{*}\eta\right)\right].\end{align}
Upon inserting  Eq. \eqref{nt11} into \eqref{nt09} and then
carrying out the resulting integrations, we get
\begin{align}\label{nt12}Q(\alpha,\beta,t)&={1\over\pi^{2}\cosh r\cosh^{2}\lambda t}\exp
\left[-\alpha^{*}\alpha-\beta^{*}\beta\right.\notag\\&\left.+(\alpha^{*}\beta^{*}+\alpha\beta)
\tanh\lambda t\right.\notag\\&\left.-{\tanh
r\over2\cosh^{2}\lambda
t}\left(\alpha^{*^{2}}+\alpha^{2}\right)\right].\end{align} We
notice that $\alpha$ and $\beta$ represent the cavity radiation
and atomic properties, respectively. Next, integrating over atomic
variables shows that
\begin{align}\label{nt13}Q(\alpha,t)&={1\over\pi\cosh r\cosh^{2}\lambda t}
\exp \left[-{1\over\cosh^{2}\lambda
t}\right.\notag\\&\times\left.\left[\alpha^{*}\alpha+{\tanh
r\over2}\left(\alpha^{*^{2}}+\alpha^{2}\right)\right]\right].\end{align}
Since the Q-function \eqref{nt12} is normalized, we realize that
Eq. \eqref{nt13} entirely corresponds to the cavity radiation.
Moreover, it is not difficult to note that in the absence of the
coupling of the atoms with the radiation, $\lambda=0$,
\begin{align}\label{nt14}Q(\alpha,t)&={1\over\pi\cosh r}
\exp \left[-\alpha^{*}\alpha-{\tanh
r\over2}\left(\alpha^{*^{2}}+\alpha^{2}\right)\right],\end{align}
which is the Q-function of the squeezed radiation alone, but when
there is no squeezed vacuum radiation, $r=0$, we  see that
\begin{align}\label{nt15}Q(\alpha,t)&={1\over\pi(1+\sinh^{2}\lambda t)}
\exp \left[-{\alpha^{*}\alpha\over(1+\sinh^{2}\lambda
t)}\right],\end{align} which  represents the radiation that would
have been emitted spontaneously from two-level atoms in
high-fitness cavity. It is known for long that such a Q-function
is associated with a radiation in a chaotic state with a mean
photon number of $sinh^{2}\lambda t$.

On the other hand, carrying out the integration over atomic
variables indicates that
\begin{align}\label{nt16}Q(\beta,t)&={1\over\pi\cosh r}
\left[{1\over\cosh^{4}\lambda t-\tanh^{2}r}
\right]^{1\over2}\notag\\&\times \exp \left[-{1
\over\cosh^{4}\lambda
t-\tanh^{2}r}\right.\notag\\&\times\left.\left[\beta^{*}\beta\big(\cosh^{2}\lambda
t-\tanh^{2} r\big) \right.\right.\notag\\&\left.\left.+{\tanh
r\sinh^{2}\lambda t
\over2}\left(\beta^{*^{2}}+\beta^{2}\right)\right]\right].\end{align}
In view of the previous discussion this Q-function describes the
dynamical evolution of the atomic properties in relation to the
population in the lower level. We notice that Eq. \eqref{nt16}
reduces for $r=0$ to
\begin{align}\label{nt17}Q(\beta,t)&={1\over\pi(1+\sinh^{2}\lambda t)}
\exp \left[-{\beta^{*}\beta\over(1+\sinh^{2}\lambda
t)}\right].\end{align} It is possible to note that Eq.
\eqref{nt17} stands for a quantum system in a chaotic state in
which the pertinent mean photon number is $sinh^{2}\lambda t$.
However, it is not difficult to observe that for $r=0$ and
$\lambda=0$ the radiation in the cavity is in a coherent state
(vacuum). Moreover, comparing Eqs. \eqref{nt15} and \eqref{nt17}
reveals that the radiation in the cavity and atomic properties in
the absence of the squeezed radiation are represented by the same
Q-function, since the source of the cavity radiation is the
spontaneous emission.

\section{Quadrature variances and mean photon number}

The squeezing properties of a single-mode radiation can be
described in terms of the quadrature operators
\begin{align}\label{nt18}\hat{a}_{+}=\hat{a}+\hat{a}^{\dagger}\end{align}
and
\begin{align}\label{nt19}\hat{a}_{-}=i(\hat{a}^{\dagger}-\hat{a}).\end{align}
The variances of these operators can be put in the anti-normal
order in the form
\begin{align}\label{nt20}\Delta a^{2}_{\pm}&=2\langle\hat{a}\hat{a}^{\dagger}\rangle
\pm\langle\hat{a}^{\dagger^{2}}\rangle\pm\langle\hat{a}^{2}\rangle
\mp\langle\hat{a}^{\dagger}\rangle^{2}\mp\langle\hat{a}\rangle^{2}\notag\\&
-2\langle\hat{a}^{\dagger}\rangle\langle\hat{a}\rangle-1,\end{align}
so that we readily employ the Q-function to obtain the expectation
values of different moments involved in Eq. \eqref{nt20}. To this
end, one can show  using Eq. \eqref{nt13} that
\begin{align}\label{nt21}\langle\hat{a}\hat{a}^{\dagger}\rangle=
\cosh^{2}\lambda t\cosh^{2}r,\end{align} as a result of which the
mean photon number of the cavity radiation turns out to be
\begin{align}\label{nt21a}\langle\hat{a}^{\dagger}\hat{a}\rangle=
\cosh^{2}\lambda t\cosh^{2}r-1.\end{align}
\begin{center}
\begin{figure}[hbt]
\centerline{\includegraphics [height=6cm,angle=0]{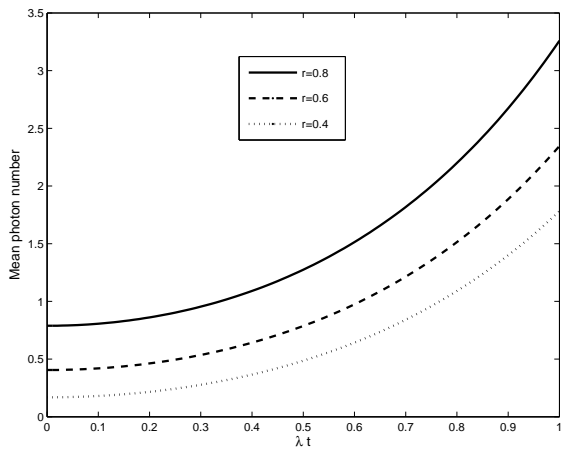}}
\caption {Plots of the mean photon number of the cavity radiation
for $\lambda=0.5$ and different values of $r$. }
\end{figure}
\end{center}

It is clearly indicated in Fig. 1 as well as Eq. \eqref{nt21a}
that the mean of the photon number of the cavity radiation
increases with the squeeze parameter and interaction time. We
realize that as time progresses much of the atoms initially in the
upper level would have a chance to decay to the lower level and
subsequently emit a radiation. We also observe that a
significantly intense light can be generated from this system
provided that the atoms are allowed to stay in the cavity for
sufficiently long period of time.

Moreover, applying Eq. \eqref{nt13} once again, it is possible to
verify that
\begin{align}\label{nt22}\langle\hat{a}^{\dagger^{2}}\rangle=
\langle\hat{a}^{2}\rangle=-\sinh r\cosh^{2}\lambda t\cosh
r,\end{align}
\begin{align}\label{nt23}\langle\hat{a}\rangle=
\langle\hat{a}^{\dagger}\rangle=0.\end{align} Hence on the basis
of Eqs. \eqref{nt20}, \eqref{nt21a}, \eqref{nt22}, and
\eqref{nt23}, we arrive at
\begin{align}\label{nt24}\Delta a^{2}_{\pm}&=2\cosh^{2}\lambda
t\cosh r\big(\cosh r\mp\sinh r\big)-1.\end{align}
\begin{center}
\begin{figure}[hbt]
\centerline{\includegraphics [height=6cm,angle=0]{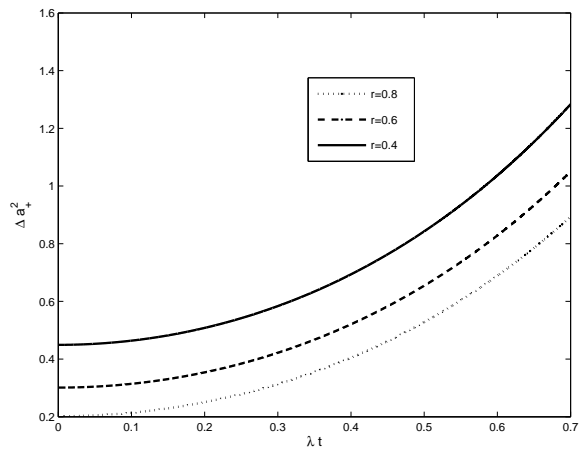}}
\caption {Plots of the plus quadrature variance of the cavity
radiation for $\lambda=0.5$ and different values of $r$. }
\end{figure}
\end{center}

It is not difficult to notice from Eq. \eqref{nt24} for $r=0$ that
the squeezing does not exist, which indicates that the squeezing
in the radiation is directly attributed to the squeezed radiation,
since the squeezing of the cavity radiation is found to be maximum
at $t=0$, that is, before the interaction with the atom is
initiated. As clearly shown in Fig. 2 the degree of squeezing of
the cavity radiation of an ensemble of two-level atoms initially
prepared to be in the upper level and placed in a squeezed vacuum
increases with the squeeze parameter, but deceases with the
interaction time. As we have discussed earlier the more the atoms
are allowed to stay in the cavity, the more probable that each
atom undergoes spontaneous emission. On the other hand, as
previously reported by various authors
\cite{pra211297,pra211573,pra241460} and as well shown in Sec. V,
the spontaneously emitted radiation from an ensemble of two-level
atoms exhibits chaotic nature. As a result, unfortunately, the
interaction of the two-level atoms with the squeezed radiation in
the cavity destroys the existing squeezing. The story would have
been different, had the atomic coherence would have been induced
either by initially preparing the atoms in the superimposed state
of the upper and lower levels or coupling the two levels
externally by a coherent light, since the induced atomic coherence
initiates additional correlation.

\section{Atom-radiation Entanglement}

One of the most interesting and intriguing ideas associated with a
composite quantum system is entanglement. A pair of particles is,
basically, taken to be entangled in quantum theory if its states
cannot be expressed as a product of the states of its individual
constituents. The preparation and manipulation of these entangled
states that have nonclassical and nonlocal properties may lead to
a better understanding of the basic quantum phenomena. It is in
this conviction that we seek to consider the entanglement of the
cavity radiation and atomic variables in this Section. In other
words, it is a well-established fact that a quantum system is said
to be entangled, if it is not separable. That is, if the density
operator for the combined state cannot be expressed as a
combination of the product density operators of the constituents,
\begin{align}\label{nt25}\hat{\rho}\ne\sum_{j}\hat{\rho}_{j}^{(1)}\bigotimes\hat{\rho}_{j}^{(2)}.\end{align}
It is now available in literature that entangled continuous
variable states can be expressed as a co-eigenstate of a pair of
EPR-type operators
 such as $\hat{X}_{a}-\hat{X}_{b}$ and
$\hat{P}_{a}+\hat{P}_{b}$ \cite{pra74}. The total variance of
these two operators reduces to zero for maximally entangled
continuous variable states. Since the cavity radiation and atomic
properties are represented in equal footing with boson operators
and the Q-function \eqref{nt12} is Gaussian in nature, it is
possible to employ the criterion set by Duan {\it{et al.}}
\cite{duan} earlier. According to this criterion, quantum states
of the composite system are entangled provided that the sum of the
variances of a pair of EPR-like operators
\begin{align}\label{nt26}\hat{u}=\hat{X}_{a}-\hat{X}_{b},\end{align}
\begin{align}\label{nt27}\hat{v}=\hat{P}_{a}+\hat{P}_{b},\end{align}
where
$\hat{X}_{a}={1\over\sqrt{2}}\big(\hat{a}+\hat{a}^{\dagger}\big)$,
$
\hat{X}_{b}={1\over\sqrt{2}}\big(\hat{b}+\hat{b}^{\dagger}\big)$,
\newline$
\hat{P}_{a}={i\over\sqrt{2}}\big(\hat{a}^{\dagger}-\hat{a}\big)$,
and
$\hat{P}_{b}={i\over\sqrt{2}}\big(\hat{b}^{\dagger}-\hat{b}\big)$,
are quadrature operators for mode $a$ and $b$, satisfy
\begin{align}\label{nt28}\Delta u^{2}+\Delta v^{2}<2,\end{align}
in which
\begin{align}\label{nt29}\Delta u^{2}+\Delta v^{2}&=2\langle\hat{a}^{\dagger}\hat{a}\rangle+
2\langle\hat{b}^{\dagger}\hat{b}\rangle+2\langle\hat{a}\rangle\langle\hat{b}\rangle+
2\langle\hat{a}^{\dagger}\rangle\langle\hat{b}^{\dagger}\rangle\notag\\&-
2\langle\hat{a}\hat{b}\rangle-2\langle\hat{a}^{\dagger}\hat{b}^{\dagger}\rangle
-2\langle\hat{a}^{\dagger}\rangle\langle\hat{a}\rangle-
2\langle\hat{b}^{\dagger}\rangle\langle\hat{b}\rangle
+2.\end{align}

 In the following, we determine the remaining correlation functions involved in Eq. \eqref{nt29}.
 To this end, with the aid of Eq. \eqref{nt16} it is possible to see that
\begin{align}\label{nt30}\langle\hat{b}^{\dagger}\hat{b}\rangle=
\cosh^{2}r\sinh^{2}\lambda t,\end{align} which represents the
atomic population in the lower level. We know that each atom emits
a radiation while decaying from the upper to lower level, in which
Eq. \eqref{nt30} can be interpreted as the mean photon number of
the emitted radiation. We notice that just like the mean number of
the cavity radiation, the mean number of the emitted photons also
increases with the squeeze parameter and interaction time. In
addition, it is not difficult to see that
\begin{align}\label{nt31}\langle\hat{a}^{\dagger}\hat{a}\rangle-
\langle\hat{b}^{\dagger}\hat{b}\rangle=\sinh^{2}r.\end{align} This
corresponds to the mean photon number of the squeezed radiation in
the cavity, since upon subtracting the emitted photons from the
cavity radiation one remains with the squeezed radiation in the
cavity. In the same way, it is possible to readily verify that
\begin{align}\label{nt32}\langle\hat{b}^{\dagger^{2}}\rangle=
\langle\hat{b}^{2}\rangle= \tanh r\cosh^{2}r\sinh^{2}\lambda
t,\end{align}
\begin{align}\label{nt33}\langle\hat{b}\rangle=
\langle\hat{b}^{\dagger}\rangle=0.\end{align} On the other hand,
the correlation between the states of the cavity radiation and
atoms is found using Eq. \eqref{nt12} to be
\begin{align}\label{nt34}\langle\hat{a}\hat{b}\rangle=
\langle\hat{a}^{\dagger}\hat{b}^{\dagger}\rangle={\sinh\lambda
t\cosh\lambda t\over\cosh r(\cosh^{2}\lambda
t-\tanh^{2}r)^{3/2}}.\end{align} Therefore, on account of Eqs.
\eqref{nt21a}, \eqref{nt23}, \eqref{nt29}, \eqref{nt30},
\eqref{nt33}, and \eqref{nt34}, we reach at
\begin{align}\label{nt36}\Delta u^{2}+\Delta
v^{2}&=2\cosh^{2}r\big(\cosh^{2}\lambda t+\sinh^{2}\lambda
t\big)\notag\\&-{4\sinh\lambda t\cosh\lambda t\over\cosh
r(\cosh^{2}\lambda t-\tanh^{2}r)^{3/2}}.\end{align}
\begin{center}
\begin{figure}[hbt]
\centerline{\includegraphics [height=6cm,angle=0]{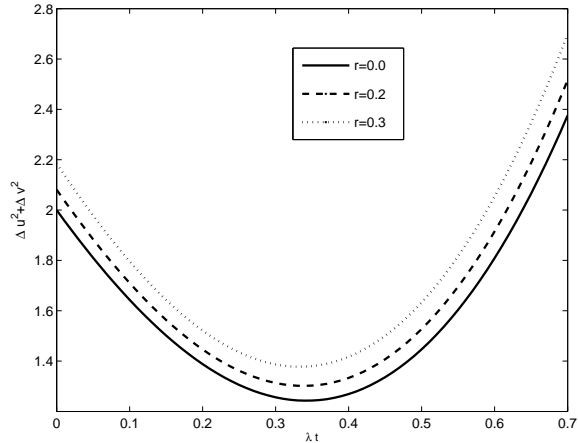}}
\caption {Plots of the sum of the variance of the EPR-type
operators of the cavity radiation for $\lambda=0.5$ and different
values of $r$. }
\end{figure}
\end{center}

It is clearly indicated in Fig. 3 that there can be a strong
atom-radiation entanglement when an ensemble of two-level atoms
initially prepared to be in the upper level are placed in a cavity
containing a squeezed radiation. We see that the entanglement
decreases with the squeeze parameter and consequently a strong
entanglement is observed in the absence of the squeezing in the
cavity, contrary to the usual expectation. This is mainly related
to the phase competition in the quadrature fluctuations. As can
readily be inferred from Eqs. \eqref{nt20} and \eqref{nt29} the
correlations leading to the squeezing and entanglement are
different, so is the contribution of the squeezed radiation
towards establishing entanglement and squeezing. This in turn
corresponds to the fact that while the correlation between similar
states results squeezing, but it is the correlation among
different states alone which is responsible for entanglement. It
is due to this underlying physical mechanism that though the
squeezed radiation is directly accountable for the squeezing of
the cavity radiation, it significantly destroys the atom-radiation
entanglement. We also observe that the entanglement is found to be
strong for a particular time corresponding to the coupling
constant and squeeze parameter. Moreover, the maximum entanglement
occurs for smaller time when larger values of the squeeze
parameter is used provided that the coupling constant is fixed. A
difference between entanglement and squeezing properties is also
observed in a down conversion process, where a correlation among
similar states of the radiation contributes to the squeezing
\cite{sintpra}. However, for nondegenerate three-level scheme it
is found that there is a direct relation between similar states in
correlated emission laser is zero due to the assumption that the
atoms are taken to be independent \cite{pra74}.

\section{Photon number distribution}

The photon number distribution for a single-mode cavity radiation
can be expressed in terms of the pertinent Q-function as
\begin{align}\label{nt37}P(n,t)={\pi\over n!}{\partial^{n}\over\partial\alpha^{n}}
{\partial^{n}\over\partial\alpha^{*^{n}}}Q(\alpha,\alpha^{*},t)|_{\alpha^{*}=\alpha=0},\end{align}
in which inserting Eq. \eqref{nt13} into \eqref{nt37} and then
carrying out the resulting differentiation reveals that
\begin{align}\label{nt38}P(n,t)&={1\over n!\cosh^{2}\lambda t\cosh r}\sum_{j,l,m=0}^{\infty}
{b^{j}c^{l+m}\over j!l!m!}
{(2l+j)!\over(2l+j-n)!}\notag\\&\times{(2m+j)!\over(2m+j-n)!}\alpha^{2l+j-n}\alpha^{*^{(2m+j-n)}}
|_{\alpha^{*}=\alpha=0},\end{align} where $b=\tanh^{2}\lambda t$
and $c=-{\tanh r\over2\cosh^{2}\lambda t}$.  Now, with the aid of
the condition $\alpha^{*}=\alpha=0$ and the fact that factorials
are defined for nonnegative integers, we obtain
\begin{align}\label{nt39}P(n,t)&={n!\over\cosh^{2}\lambda t\cosh r}\sum_{j=0}^{n}
{\tanh^{2j}\lambda t\big({\tanh r\over\cosh^{2}\lambda
t}\big)^{n-j}\over2^{n-j}j!\left[\left({n-j\over2}\right)!\right]^{2}},\end{align}
since $n-j$ should be even. Following a straight forward
calculations Eq. \eqref{nt39} is found to reduce, in the absence
of the squeezed radiation, to
\begin{align}\label{nt40}P(n,t)&={(\sinh^{2}\lambda t)^{n}\over(1+\sinh^{2}\lambda
t)^{n+1}}.\end{align} This is the photon number distribution of
the radiation that would have been emitted spontaneously from
two-level atoms. We see from Eq. \eqref{nt21a} that for $r=0$,
$\langle\hat{a}^{\dagger}\hat{a}\rangle=\sinh^{2}\lambda t$.
Therefore, Eq. \eqref{nt40} indicates that the spontaneously
emitted radiation from an ensemble of two-level atoms prepared
initially in the upper level and placed in the cavity initially
maintained at a vacuum state is in a chaotic state, which is
consistent with earlier reports \cite{pra211297,
pra241460,pra211573}.

\section{Conclusion}

The analysis of the interaction of an ensemble of two-level atoms
initially prepared in the upper level and placed in a cavity
containing a squeezed radiation with a cavity radiation is
presented. The squeezing as well as the statistical properties of
the cavity radiation is thoroughly studied. On the basis of the
fact that the spontaneously emitted radiation is found to be in
the chaotic state, the squeezing of the cavity radiation decreases
with time of interaction. It is a well established fact that the
longer the time of interaction, the more would be the intensity of
the radiation in a cavity. But, unfortunately, the cavity
radiation completely losses its squeezing properties in case when
intense light can be generated. We, hence, cannot see the utility
of this system as a source of a squeezed light as usually expected
\cite{josab41490}.

The photons emitted spontaneously by an ensemble of two-level
atoms are uncorrelated. However, the result in Sec. IV clearly
shows that there is a significant correlation between the emitted
photons and the atoms though the exchange of the spontaneously
emitted radiation. This correlation is sufficiently strong to
warrant atom-radiation entanglement. We find that this
entanglement depends on the time of interaction. The entanglement
becomes large at a particular time that depends on the squeeze
parameter and coupling constant. We also observe that the
squeezing of the radiation destroys the atom-radiation
entanglement contrary to our expectation that due to the
correlation initiated by squeezing the entanglement would have
increased. We believe that this effect of the squeezed radiation
may be related to the phase competition in the quadrature
fluctuations. The controversy in the fact that the squeezed
radiation is completely responsible for cavity squeezing although
it degrades entanglement is associated with the difference in
correlation that leads to these quantum features of the radiation.
The same effect of the squeezed radiation is recently seen in
harmonic entanglement of a driven degenerate parametric oscillator
when the cavity is coupled to the squeezed vacuum reservoir
\cite{sintpra}.

{\bf{ Acknowledgments}}

 I thank  Dr. Fesseha
Kassahun for introducing me to the method of evaluating the
coherent-state propagator.

\end{document}